\documentclass[a4paper,conference]{IEEEtran}
\IEEEoverridecommandlockouts

\usepackage{cite}
\usepackage{amsmath,amssymb,amsfonts}
\usepackage{multirow}
\usepackage{algorithmic}
\usepackage{graphicx}
\usepackage{textcomp}
\usepackage{xcolor}
\def\BibTeX{{\rm B\kern-.05em{\sc i\kern-.025em b}\kern-.08em
    T\kern-.1667em\lower.7ex\hbox{E}\kern-.125emX}}
\begin{document}

\title{Analytical Approach to Wave Scattering \\in Waveguide Junction \\with Conducting Cylindrical Posts\\
\thanks{This work was supported by the Gdansk University of Technology through the Excellence Initiative–Research University within the Argentum Triggering Research Grants under Grant no DEC-19/1/2023/IDUB/I3b/Ag.}
}

\author{\IEEEauthorblockN{Malgorzata Warecka, Rafal Lech, Piotr Kowalczyk}
\IEEEauthorblockA{\textit{Department  of  Microwave  and  Antenna  Engineering,   Faculty   of   Electronics,   Telecommunications   and   Informatics} \\
\textit{Gdansk  University  of  Technology}\\
Gdansk, Poland \\
malgorzata.warecka@pg.edu.pl, rafal.lech@pg.edu.pl, pio.kow@gmail.com}
}

\maketitle

\begin{abstract}
A new approach for analyzing waveguide junctions containing conductive cylindrical objects is proposed. The algorithm is based on mode matching technique using local projection functions, which improves the numerical conditioning of the problem. Moreover, the approach enhances computational efficiency by reducing the boundary where the numerical integration is required. Both convergence and accuracy of the method were tested by using several examples, including multi-section microwave filters, and validated through comparison with results obtained using the finite element method implemented in a commercial full-wave simulator.

\end{abstract}


\section{Introduction}
Waveguide technology has been used for decades in terrestrial and satellite communications, radar systems, and other high-power applications. Recent years have seen increased interest and rapid development in space and satellite technologies. Consequently, the demands on tools that enable fast and efficient simulation of structures used in these systems have also been growing \cite{CruzJM25,baranowskiMWCL22,sorockiMTT24,huescarMTT2024}.

Commonly used discrete methods are highly versatile, which is the reason for their widespread implementation in commercial full-wave simulators \cite{inventsim,ansys,cst}. However, the versatility of these methods comes at the cost of limited efficiency. As a consequence, software dedicated to specific structures or geometries is constantly being developed. In such cases, a hybrid approach \cite{kusiekAP08,wareckaMTT20,RezaiesarlakJRFMCAE2011} or sometimes even analytical methods \cite{polewskiMTT04, lechMTT07} can be used.

This paper analyzes the analytical description of the electromagnetic wave scattering phenomenon in a rectangular waveguide junction containing cylindrical conducting objects. Similar structures have already been analyzed in the literature \cite{kusiekAP08,polewskiMTT04, lechMTT07}. In all these cases, the analysis involved domain decomposition and introduced a hypothetical cylindrical region containing the investigated objects. The cylinder filled completely the junction (the diameter of the cylindrical region was equal to the width of the waveguide) and was connected to waveguide ports in which the incident and reflected waves were defined.
The electromagnetic fields in the waveguide sections, typically expressed in Cartesian coordinates, were matched with the field in the cylindrical region, expressed as cylindrical functions. The projection used to fulfill the continuity conditions requires integration over the entire junction cross-section. In such cases, the integrals must have been evaluated numerically due to their complexity. Furthermore, the number of cylindrical functions within the junction and modes in the waveguides were closely related, leading to limitations on the number of modes, thus limiting the applicability of the method.

In this paper an alternative approach is proposed, in which the junction region does not require the introduction of a hypothetical cylindrical region, and the numerical integration resulting from the projection of fields during matching is limited to the boundary of the scattering object. We use local basis functions known from one-dimensional finite element method (FEM) for this projection, since projecting global functions can lead to poor numerical conditioning. In this article, we analyze only conductive cylindrical objects and demonstrate the validity of the developed algorithm by comparing the obtained results with those obtained from the full-wave electromagnetic simulator InventSim \cite{inventsim}. It should be noted that the presented technique can also be adapted to analyze non-conductive objects with geometries other than cylindrical, described, for example, by an impedance matrix defined on the cylindrical surface surrounding the object. This topic will be addressed in our future work.

\section{Formulation}
\begin{figure}[htbp]
\centerline{\includegraphics[width=0.9\columnwidth]{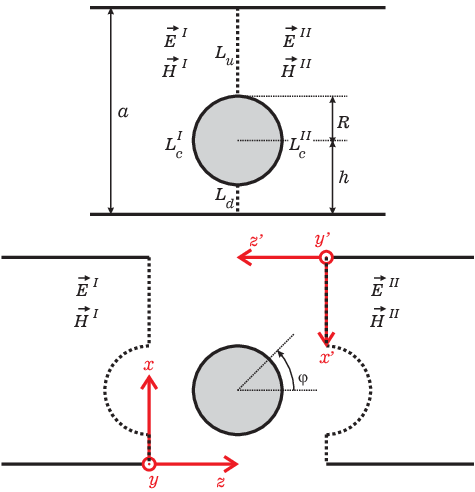}}
\caption{Geometry of the considered waveguide junction}
\label{fig_structure}
\end{figure}

The analysis concerns a conductive cylindrical post located in a rectangular waveguide junction.
It is assumed that the cylinder extends over the full height $b$ of the waveguide.
Fig.~\ref{fig_structure} shows geometry of the structure where a center of the post of radius $R$ is located at a distance $h$ from the waveguide wall.  Further analysis addresses only the TE$_{m0}$ modes propagating in the junction.

To determine the scattering matrix of the section, it is necessary to uniquely define the structure's ports. We assume that the junction length is zero, and conventionally, port $\mathrm{I}$ is located at $z=0$ plane, while port $\mathrm{II}$ at $z'=0$ plane. The fact that the cylindrical object is not entirely contained within the junction (section length equals zero) does not affect the validity of the analysis as shown for larger hypothetical cylinders \cite{kusiekAP08,polewski}. The analysis requires defining the electric and magnetic fields in the cross-sections for both ports:
\begin{equation}
    E^{I}_y(x,z) = \sum^{M}_{m=1}p_m G_m (a^{I}_m e^{-\gamma_m z}+b^{I}_m e^{\gamma_m z})\sin (p_m x)
\end{equation}
\begin{equation}
    E^{II}_y(x',z') = \sum^{M}_{m=1}p_m G_m (a^{II}_m e^{-\gamma_m z'}+b^{II}_m e^{\gamma_m z'})\sin (p_m x')
\end{equation}
\begin{equation}
 H^{I}_x(x,z) = \sum^{M}_{m=1} p_m\frac{G_m}{Z_m}(b^{I}_m e^{\gamma_m z}-a^{I}_m e^{-\gamma_m z})\sin (p_m x)  
\end{equation}
\begin{equation}
  H^{II}_x(x',z') = \sum^{M}_{m=1} p_m\frac{G_m}{Z_m}(b^{II}_m e^{\gamma_m z'}-a^{II}_m e^{-\gamma_m z'})\sin (p_m x')  
\end{equation}
where the parameters $p_m=m\pi/a$, $\gamma_m=\sqrt{p^2_m-\omega^2\mu\varepsilon}$, $G_m = \sqrt{(2\mathrm{j}\omega\mu)/(\gamma_m ab p^2_m)}$, $Z_m=\mathrm{j}\omega\mu/\gamma_m$ depend on the waveguide medium. The coefficients $a^{I}_m$, $b^{I}_m$, $a^{II}_m$ and $b^{II}_m$ represent the incoming and outgoing signals at port $\mathrm{I}$ and $\mathrm{II}$, respectively.

The tangential components of the electric and magnetic fields in both ports must be continues on the segments $L_d=\{x\in[0,h-R]\wedge z=0\}$ and $L_u=\{x\in[h+R,a]\wedge z=0\}$. In the case of boundaries with a conductor $L_c^I=\{x(\varphi)=h+R\sin(\varphi)\wedge z(\varphi)=R\cos(\varphi), \textrm{ where } \varphi\in[\pi/2,3\pi/2]\}$ and $L_c^{II}=\{x'(\varphi)=a-h-R\sin(\varphi)\wedge z'(\varphi)=-R\cos(\varphi), \textrm{ where } \varphi\in[-\pi/2,\pi/2]\}$ only the tangential component of the electric field is equal to zero:
\begin{equation}
    \begin{cases}
    H^{I}_x(x,0)= H^{II}_{x'}(x',0)\Big|_{L_d \cup L_u}\\
      E^{I}_y(x,0) = E^{II}_{y'}(x',0)\Big|_{L_d \cup L_u}\\
      E^{I}_y(x,z)\Big|_{L_{c1}} = 0\\ 
      E^{II}_{y'}(x',z')\Big|_{L_{c2}} = 0
    \end{cases}     
    \label{eq_bnd}
\end{equation}
The relations between the coordinates in the local coordinate systems used to describe the fields in each port are very simple: $x' = a-x$, $y'=y$, and $z'=-z$. The orientation of the axes of the systems is as follows: $\vec{i}_{x'}=-\vec{i}_x$, $\vec{i}_{y'}=\vec{i}_y$, and $\vec{i}_{z'}=-\vec{i}_z$.

\begin{figure}[htbp]
\centerline{\includegraphics[width=0.9\columnwidth]{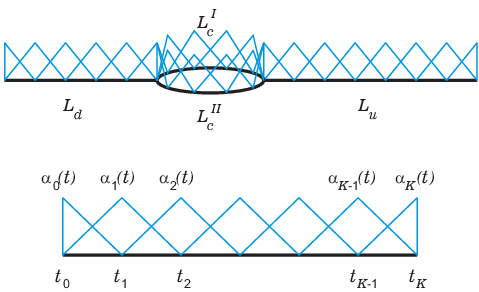}}
\caption{Visualization of local basis functions on boundaries}
\label{fig_bfun_total}
\end{figure}

Fulfilling the continuity conditions \eqref{eq_bnd} is traditionally achieved by projection using an appropriate set of functions. Unfortunately, global functions defined along the entire junction cross-section ($L_d \cup L_c^{I} \cup L_u$) or ($L_d \cup L_c^{II} \cup L_u$) are poorly conditioned numerically; as the number of functions increases, their linear independence deteriorates, leading to an ill-conditioned system. In the presented approach, we propose the use of local basis functions used in the one-dimensional (first-order) FEM. Their restricted domain improves numerical conditioning, and additionally, most integrations can be performed analytically. Each of the boundary segments ($L_d$, $L_c^{I}$, $L_c^{II}$ and $L_u$) is uniformly discretized into subdomains $\Omega_k$, on which the following local functions are defined (see Fig.~\ref{fig_bfun_total}). Each of the local functions $\alpha_k (t)$ is nonzero only on its assigned subdomain $\Omega_k$. The formula is as follows 
\begin{eqnarray}
     &&\!\!\!\!\!\!\!\!\!\alpha_k (t) = 1 - \frac{2|t-t_k|}{t_{k+1}-t_{k-1}}, \quad t \in [t_{k-1},t_{k+1}]=\Omega_k
\end{eqnarray}
for internal regions of the domain where $k = 1,\dotsc, K-1$, while for the edges
\begin{eqnarray}
    &&\!\!\!\!\!\!\!\!\!\!\!\!\alpha_0 (t) = 1 - \frac{t-t_0}{t_1-t_0}, \quad\qquad t \in [t_0,t_1]=\Omega_0\\
    &&\!\!\!\!\!\!\!\!\!\!\!\!\alpha_K (t) = 1 + \frac{t-t_K}{t_{K}-t_{K-1}}, \quad t \in [t_{K-1},t_{K}]=\Omega_K
\end{eqnarray}
where $t_k=t_0+k(t_k-t_0)/K$. The range and exact form of the function for each of the limit curves are specified in Table \ref{tab1}.

\begin{table}[htbp]
\caption{Local basis function parameters}
\begin{center}
\begin{tabular}{|c||c|c|c|c|c|}
\hline
&$t$ & $t_0$ & $t_K$ & $K$ & $\Omega_k$\\
\hline
\hline
& & & & &\\
$\alpha^d_k$& \multirow{2}{*}{$x$}  & $0$ & $h-R$ & $K_d$ & $\Omega_k^d$ \\
$\alpha^u_k$& & $h+R$ & $a$ & $K_u$ & $\Omega_k^u$\\
& & & & &\\
\hline
& & & & &\\
$\alpha^{cI}_k$& \multirow{2}{*}{$\varphi$} & $ \frac{\pi}{2}$ & $ \frac{3\pi}{2}$ & $K_{c1}$ & $\Omega_k^{cI}$\\
$\alpha^{cII}_k$& & $-\frac{\pi}{2}$ & $ \frac{\pi}{2}$ & $K_{c2}$ & $\Omega_k^{cII}$\\
& & & & &\\
\hline
\end{tabular}
\label{tab1}
\end{center}
\end{table}

Projection of the conditions \eqref{eq_bnd} using the local functions defined above leads to a system of equations in which the only unknowns are the coefficients representing the incoming and outgoing signals at the ports $\mathrm{I}$ and $\mathrm{II}$. We obtain $2K_d$ equations (for $k=1,\dots, K_d$)
\begin{eqnarray}
&&\!\!\!\!\!\!\!\!\!\!\!\!\!\!\int_{L_d}H^{I}_x(x,0)\alpha^d_k(x)\;dx =  \int_{L_d}H^{II}_{x'}(x',0)\alpha^d_k(x)\;dx\\
&&\!\!\!\!\!\!\!\!\!\!\!\!\!\!\int_{L_d}E^{I}_y(x,0)\alpha^d_k(x)\;dx =  \int_{L_d}E^{II}_{y'}(x',0)\alpha^d_k(x)\;dx
\end{eqnarray}
for the conditions concerning segment $L_d$. Similarly, it will be $2K_u$ equations (for $k=0,\dots, K_u-1$)
\begin{eqnarray}
&&\!\!\!\!\!\!\!\!\!\!\!\!\!\!\int_{L_u}H^{I}_x(x,0)\alpha^u_k(x)\;dx =  \int_{L_u}H^{II}_{x'}(x',0)\alpha^u_k(x)\;dx \\
&&\!\!\!\!\!\!\!\!\!\!\!\!\!\!\int_{L_u}E^{I}_y(x,0)\alpha^u_k(x)\;dx =\int_{L_u}E^{II}_{y'}(x',0)\alpha^u_k(x)\;dx 
\end{eqnarray}
for segment $L_u$. The conditions defined on the cylinder wall $L_c^{I}$ and $L_c^{II}$ lead to $2(K_c+1)$ equations (for $k=0,\dots, K_c$)
\begin{eqnarray}
\int_{L_c^I}E^{I}_{y}(x(\varphi),z(\varphi))\alpha^{cI}_k(\varphi)\;d\varphi &=&0 \\
\int_{L_c^{II}}E^{II}_{y'}(x'(\varphi),z'(\varphi))\alpha^{cII}_k(\varphi)\;d\varphi &=&0
\end{eqnarray}
The resulting system of equations can be briefly written in the matrix notation
\begin{equation}
    \mathbf{L} 
   \mathbf{b}
    = \mathbf{R}
    \mathbf{a}
    \label{eq_matrixLR}
\end{equation}
where
\begin{equation}
 \mathbf{L}=
    \begin{bmatrix}
     -\mathbf{H^{xId}} & \mathbf{H^{xIId}} \\
    \mathbf{E^{yId}} & -\mathbf{E^{yIId}} \\
    -\mathbf{H^{xIu}} & \mathbf{H^{xIIu}} \\
    \mathbf{E^{yIu}} & -\mathbf{E^{yIIu}} \\
    \mathbf{E^{yIcb}} & \mathbf{0} \\
    \mathbf{0} & \mathbf{E^{yIIcb}}
    \end{bmatrix},
         \mathbf{R}=
    \begin{bmatrix}
    -\mathbf{H^{xId}} & \mathbf{H^{xIId}} \\
    -\mathbf{E^{yId}} & \mathbf{E^{yIId}} \\
    -\mathbf{H^{xIu}} & \mathbf{H^{xIIu}} \\
    -\mathbf{E^{yIu}} & \mathbf{E^{yIIu}} \\
    -\mathbf{E^{yIca}} & \mathbf{0} \\
    \mathbf{0} & -\mathbf{E^{yIIca}}
    \end{bmatrix}
\end{equation}
whereby
\begin{eqnarray}
    \mathbf{a}= 
    \begin{bmatrix}
    a^{I}_1,a^{I}_2,\ldots,a^{I}_M,a^{II}_1,a^{II}_2,\ldots,a^{II}_M
    \end{bmatrix}^T
    \\
    \mathbf{b}= 
    \begin{bmatrix}
    b^{I}_1,b^{I}_2,\ldots,b^{I}_M,b^{II}_1,b^{II}_2,\ldots,b^{II}_M
    \end{bmatrix}^T
\end{eqnarray}
while the submatrices are defined as follows:
\begin{eqnarray}
    &&\!\!\!\!\!\!\!\!\!\!\!\!\!\!\mathbf{H^{xId}}_{k,m}=-\int_{\Omega_k^d} p_m\frac{G_m}{Z_m}\sin (p_m x)  \alpha_k^d(x)\;dx\label{eq:19}\\
    &&\!\!\!\!\!\!\!\!\!\!\!\!\!\!\mathbf{H^{xIId}}_{k,m}=-\int_{\Omega_k^d} p_m\frac{G_m}{Z_m}\sin (p_m x')  \alpha_k^d(x)\;dx\\
    &&\!\!\!\!\!\!\!\!\!\!\!\!\!\!\mathbf{H^{xIu}}_{k,m}=-\int_{\Omega_k^u} p_m\frac{G_m}{Z_m}\sin (p_m x)  \alpha_k^u(x)\;dx\\
    &&\!\!\!\!\!\!\!\!\!\!\!\!\!\!\mathbf{H^{xIIu}}_{k,m}=-\int_{\Omega_k^u} p_m\frac{G_m}{Z_m}\sin (p_m x')  \alpha_k^u(x)\;dx\\
    &&\!\!\!\!\!\!\!\!\!\!\!\!\!\!\mathbf{E^{yId}}_{k,m}=\int_{\Omega_k^d} p_m G_m \sin (p_m x)  \alpha_k^d(x)\;dx\\
     &&\!\!\!\!\!\!\!\!\!\!\!\!\!\!\mathbf{E^{yIId}}_{k,m}=\int_{\Omega_k^d} p_m G_m \sin (p_m x')  \alpha_k^d(x)\;dx\\
     &&\!\!\!\!\!\!\!\!\!\!\!\!\!\!\mathbf{E^{yIu}}_{k,m}=\int_{\Omega_k^u} p_m G_m \sin (p_m x)  \alpha_k^u(x)\;dx\\
     &&\!\!\!\!\!\!\!\!\!\!\!\!\!\!\mathbf{E^{yIIu}}_{k,m}=\int_{\Omega_k^u} p_m G_m \sin (p_m x')  \alpha_k^u(x)\;dx\label{eq:26}\\
     &&\!\!\!\!\!\!\!\!\!\!\!\!\!\!\mathbf{E^{yIcb}}_{k,m}=\nonumber\\&&\!\!\!\!\!\!\!\!\!\!\!\!\!\!\qquad=\int_{\Omega_k^{cI}} p_m G_m e^{\gamma_m z}\sin (p_m x)\alpha_k^{cI}(\varphi)\;d\varphi\\
     &&\!\!\!\!\!\!\!\!\!\!\!\!\!\!\mathbf{E^{yIIcb}}_{k,m}=\nonumber\\&&\!\!\!\!\!\!\!\!\!\!\!\!\!\!\qquad=\int_{\Omega_k^{cII}} p_m G_m e^{\gamma_m z'}\sin (p_m x')\alpha_k^{cII}(\varphi)\;d\varphi\label{eq:27}\\
     &&\!\!\!\!\!\!\!\!\!\!\!\!\!\!\mathbf{E^{yIca}}_{k,m}=\nonumber\\&&\!\!\!\!\!\!\!\!\!\!\!\!\!\!\qquad=\int_{\Omega_k^{cI}} p_m G_m e^{-\gamma_m z}\sin (p_m x)\alpha_k^{cI}(\varphi)\;d\varphi\\
     &&\!\!\!\!\!\!\!\!\!\!\!\!\!\!\mathbf{E^{yIIca}}_{k,m}=\nonumber\\&&\!\!\!\!\!\!\!\!\!\!\!\!\!\!\qquad=\int_{\Omega_k^{cII}} p_m G_m e^{-\gamma_m z'}\sin (p_m x')\alpha_k^{cII}(\varphi)\;d\varphi\label{eq:30}
\end{eqnarray}

It should be emphasized that the integrals \eqref{eq:19}-\eqref{eq:26} can be calculated analytically, which significantly improves computational efficiency in comparison to the classical approach \cite{kusiekAP08,polewskiMTT04}.

With an appropriately selected number of modes and basis functions (the condition of definiteness of the system $M<K_u+K_d+K_c+1$ must be met) from the relation \eqref{eq_matrixLR} one can obtain an expression for the scattering matrix $\mathbf{S}$
\begin{equation}
   \mathbf{b}
    =    \mathbf{L}^{-1} \mathbf{R} \mathbf{a}=  \mathbf{S} \mathbf{a}
\end{equation}
In practice, the matrix $\mathbf{L}$ is not inverted, the resulting system of equations is overdetermined and it is solved by direct methods, such as the least squares method.

\section{Numerical results}

To verify the validity and accuracy of the developed algorithm, three waveguide structures containing conductive cylinders were analyzed. The results were compared with those obtained from the InventSim full-wave simulator \cite{inventsim} based on FEM.

\begin{figure}[htbp]
\centerline{\includegraphics[width=0.9\columnwidth]{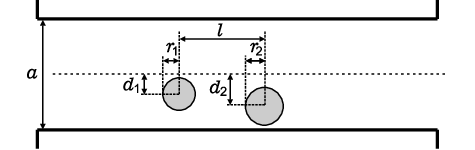}}
\caption{Section of a rectangular waveguide with two cylindrical resonators of radii $r_1$ and $r_2$, separated by a distance $l$ and offset from the waveguide axis by distances $d_1$ and $d_2$ (top view).}
\label{fig_fal2}
\end{figure}

The first structure is a WR-62 waveguide junction (width $a=15.799$ mm, height $b=7.899$ mm) containing two conducting cylindrical posts of radius $r = 2$ mm. The cylinders are offset by distances $d_1=3$ mm and $d_2=5$ mm relative to the waveguide axis and separated by a distance $l$. A geometry of the analyzed structure is shown in Fig.~\ref{fig_fal2}.

\begin{figure}[htbp]
\centerline{\includegraphics[width=0.9\columnwidth]{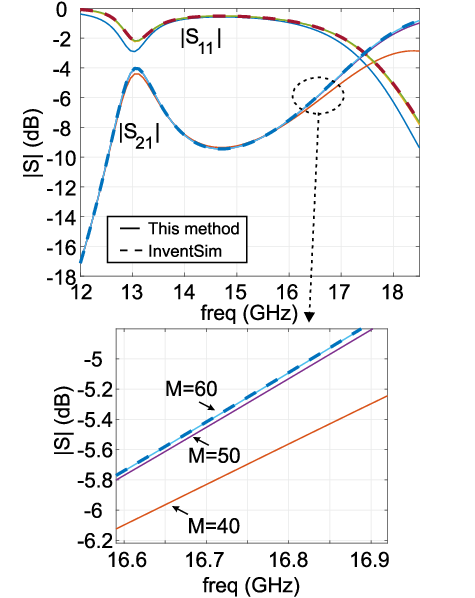}}
\caption{Frequency responses of the structure shown in Fig. \ref{fig_fal2} for the following dimensions: $r_1=r_2=2$~mm, $d_1=3$~mm, $d_2=5$~mm, $l=15$~mm, in the WR-62 waveguide.}
\label{fig_res15}
\end{figure}

\begin{figure}[htbp]
\centerline{\includegraphics[width=0.9\columnwidth]{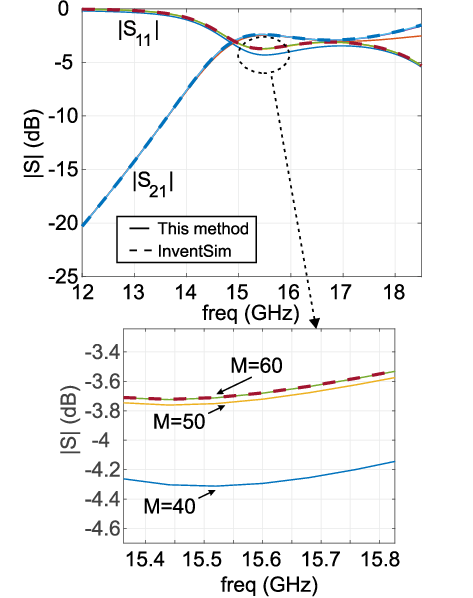}}
\caption{Frequency responses of the structure shown in Fig. \ref{fig_fal2} for the following dimensions: $r_1=r_2=2$~mm, $d_1=3$~mm, $d_2=5$~mm, $l=10$~mm, in the WR-62 waveguide.}
\label{fig_res10}
\end{figure}

\begin{figure}[htbp]
\centerline{\includegraphics[width=0.9\columnwidth]{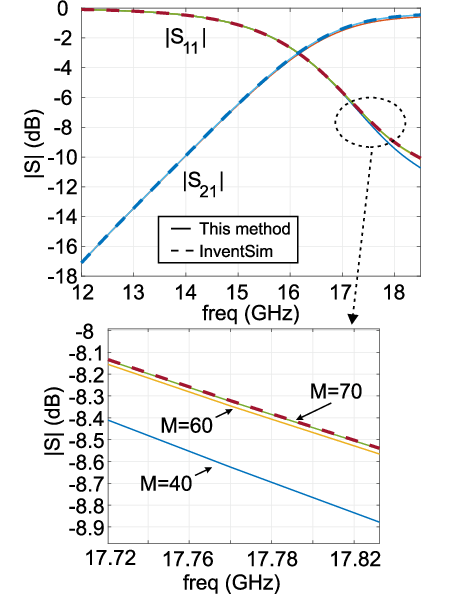}}
\caption{Frequency responses of the structure shown in Fig. \ref{fig_fal2} for the following dimensions: $r_1=r_2=2$~mm, $d_1=3$~mm, $d_2=5$~mm, $l=5$~mm, in the WR-62 waveguide.}
\label{fig_res5}
\end{figure}

Calculations were performed for various cylinder separations $l$ and different numbers of modes $M$ to examine the convergence of the method. Figs.~\ref{fig_res15}, \ref{fig_res10}, and \ref{fig_res5} show the results for cylinders separated by distances of $15$~mm, $10$~mm, and $5$~mm, respectively. 

The results show that, for distances between cylinders of $10$~mm and $15$~mm, the algorithm converges when $M = 60$, whereas for smaller separations (e.g., $5$~mm), $M = 70$ is required. This behavior results from the excitation of higher-order field components at the discontinuity (the cylindrical post) inside the waveguide. Although these modes do not propagate within the fundamental-mode frequency band, their influence must be considered when cascading waveguide sections.

\begin{figure}[htbp]
\centerline{\includegraphics[width=0.9\columnwidth]{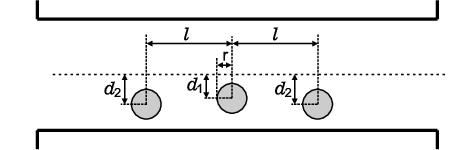}}
\caption{Structure of a three-post conducting waveguide filter (top view).}
\label{fig_fil3}
\end{figure}

\begin{figure}[htbp]
\centerline{\includegraphics[width=0.9\columnwidth]{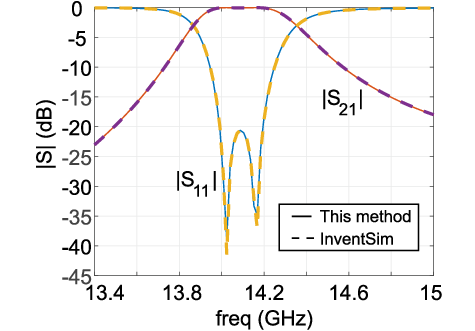}}
\caption{Frequency response of the three-post conducting waveguide filter shown in Fig.~\ref{fig_fil3}. Dimensions: $r=2$ mm, $l=14.7404$ mm, $d_1=3.4475$ mm, $d_2=1.5137$ mm, waveguide WR-62.}
\label{fig_fil3_res}
\end{figure}

\begin{figure}[htbp]
\centerline{\includegraphics[width=0.9\columnwidth]{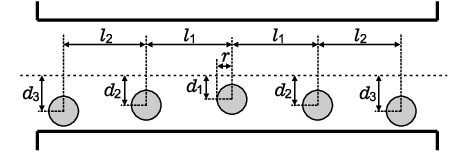}}
\caption{Structure of a five-post conducting waveguide filter (top view).}
\label{fig_fil5}
\end{figure}

\begin{figure}[htbp]
\centerline{\includegraphics[width=0.9\columnwidth]{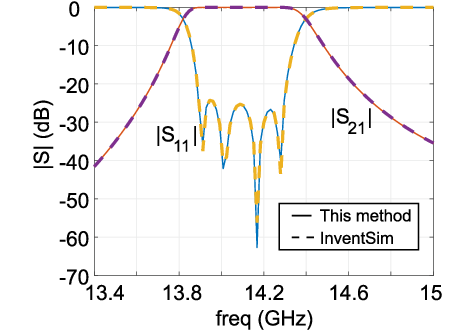}}
\caption{Frequency response of the five-post conducting waveguide filter shown in Fig.~\ref{fig_fil5}. Dimensions: $r=2$ mm, $l_1=14.1461$ mm, $l_2=15.9014$ mm, $d_1=3.9639$ mm, $d_2=1.7958$ mm, $d_3=1.3672$ mm, waveguide WR-62.}
\label{fig_fil5_res}
\end{figure}

After determining the appropriate number of modes from the convergence analysis, more complex filter structures were analyzed. The second and third structures are bandpass filters with two and four poles, respectively, with the reflection characteristics described in \cite{polewski}. The results for the two-pole filter (see Fig.~\ref{fig_fil3}) are shown in Fig.~\ref{fig_fil3_res}, while those for the four-pole filter (see Fig.~\ref{fig_fil5}) are shown in Fig.~\ref{fig_fil5_res}. The dimensions of the structures are provided in the figure captions.



In both cases, the results were compared with those obtained from the InventSim full-wave simulator, and a good agreement was obtained.

\section{Conclusion}
This paper presented an analytical formulation for the scattering of electromagnetic waves in a rectangular waveguide junction containing conductive cylindrical objects. A new field-matching approach based on local projection functions was proposed, offering improved numerical conditioning and efficiency compared to previously published methods. The algorithm was verified for convergence and validated against results obtained using a commercial full-wave electromagnetic simulator, demonstrating very good agreement.


\vspace{12pt}

\end{document}